\documentclass[intlimits,twoside,a4paper]{article}

\usepackage{amsmath,amssymb}
\usepackage{graphicx}

\usepackage[T2A]{fontenc}
\usepackage[cp1251]{inputenc}
\usepackage[eqsecnum]{cmpj2}

\issue{2014}{17}{2}{23703}
\doinumber{10.5488/CMP.17.23703}

\title[Interlevel absorption of electromagnetic waves \ldots]%
{Interlevel absorption of electromagnetic waves by nanocrystal with divalent impurity}
\author[V.I Boichuk, R.Ya. Leshko]{V.I Boichuk, R.Ya. Leshko\thanks{E-mail: leshkoroman@gmail.com}}
\address{Department of Theoretical Physics, Ivan Franko Drohobych State Pedagogical University, \\
3 Stryiska St., 82100 Drohobych, Ukraine}

\date{Received March 19, 2014, in final form May 19, 2014}
\authorcopyright{V.I Boichuk, R.Ya. Leshko, 2014}

\begin{document}

\maketitle

\begin{abstract}
The energy spectrum of central divalent impurity is calculated using the
effective mass approximation in a spherical quantum dot (QD). The dipole moment and oscillator
strength of interlevel transition is defined. The dependence of linear absorption coefficient
on the QD size and electromagnetic frequency is analyzed. The  obtained
results are compared with the results of univalent impurity.
\pacs 73.21.La, 78.20.Ci
\end{abstract}

\section{Introduction}

The semiconductor quantum dots (QDs) are widely used in opto- and nanoelectronics due to their unique properties.
Lasers, sources of light, LEDs are constructed based on nanosystems. Sources of
terahertz radiation, which are constructed based on QDs, take a special place \cite{Wu}. The
feature of terahertz radiation lies in the fact that it practically does not ionize
materials, contrary to the X-ray, and is capable of penetrating into materials. That is why this kind of
radiation is widely used in medical tomography \cite{Wang}, in security systems,
in producing high resolution images of microscopic objects \cite{Huber}. The possibilities of
developing high-speed THz communication systems are studied \cite{Piesiewicz}. The detector
of terahertz radiation was proposed based on QDs \cite{Wei1,Wei2}. Taking into consideration
that the energy of interlevel transitions responds to the terahertz range, the study of interlevel
transitions became the basis for theoretical description and prognostication of the properties of terahertz
detectors and sources.

Single-electron states in the QD which
definitely depend on the QD size, the presence of defects, especially impurities, are the basis of interlevel transition analysis.
At present, the theory of shallow hydrogenic donor impurities is widely developed in the QD. An exact solution of
Schr\"{o}dinger equation for the central impurity was derived \cite{Tkach}, the energy spectrum of
the off-central impurity was obtained using different methods in spherical \cite{Boichuk1}
and ellipsoidal \cite{Sadeghi} QDs. The cubic \cite{Rezaei1} QDs are analysed too. Since the QD can
contain several impurities, the problem regarding the QD with two impurities was solved \cite{Holovatsky1,Boichuk2}.
Based on the obtained results, the linear and nonlinear optical properties of the QD with
impurities \cite{Boichuk1,Sadeghi,Rezaei1,Boichuk2,Vahdani,Rezaei2} were calculated using the density matrix and iteration method \cite{Tang}.

Experimental data show that QDs can be doped with impurities which are divalent \cite{Korb}.
In particular, in this work it was shown that the zinc impurities penetrate the CdS QDs.
This leads to the changes of the optical properties which are connected with interband (high-energy)
and interlevel intraband (low-energy) transitions.

The above mentioned as well as the lack of a consistent theory of central divalent impurities in spherical QDs,
which could make possible the calculation of the ground and excited states, brings about the necessity to consider the divalent
impurity in a spherical QD; to determine the energy spectrum of this impurity;
to calculate interlevel transitions in the QD with divalent impurity;
to compare the obtained results with the corresponding results of  monovalent impurity.

\section{Eigenvalues and eigenfunctions}

We consider a spherical nanosize heterosystem. It consists of a nanocrystal of radius $a$ having electron
effective mass $m_1^*$, which is placed in a matrix having electron effective mass $m_2^*$. There is a
divalent impurity in the center of the QD. Let the heterosystem be made of crystals that have
the values close to dielectric permittivity. This makes it possible to introduce the average value of
dielectric permittivity $\varepsilon$. The effective-mass Hamiltonian of this system can be written as follows:
\begin{equation}\label{1}
   \hat H = \hat H_1^{} + \hat H_2^{} + \frac{{{e^2}}}{{4\pi {\varepsilon _0}\varepsilon {r_{12}}}}\, ,
\end{equation}
where
\begin{equation}\label{2}
    \hat H_i^{} =  - \frac{{{\hbar ^2}}}{2}{\nabla _i}\frac{1}{{{m^*}\left( {{r_i}}
    \right)}}{\nabla _i} + U({r_i}) - \frac{{Z{e^2}}}{{4\pi {\varepsilon _0}\varepsilon {r_i}}} = \hat H_i^{(0)} - \frac{{Z{e^2}}}{{4\pi {\varepsilon _0}\varepsilon {r_i}}}\,,
\end{equation}
$Z=2$. The potential energy caused by the heterostructure band mismatch is given by:
\begin{equation}\label{3}
    U({r_i}) = \left\{ \begin{array}{ll}
    0, & \hbox{${r_i} \leqslant a$},\\
    {U_0},& \hbox{${r_i} > a$}.
    \end{array} \right.
\end{equation}
The Schr\"{o}dinger equation with the Hamiltonian (\ref{1}) cannot be solved exactly.
Therefore, the Ritz variation method has been used herein. Since the electrons are
fermi-particles, the wave function should be antisymmetric.
The approach of \cite{Boichuk1,Boichuk3,Boic4} has been used for the chosen variation function.
Nonetheless in \cite{Boichuk3,Boic4}
there was calculated only the ground state energy of divalent impurity, and in \cite{Boichuk1}
there was calculated the energy of the ground state and the first exited states of the monovalent impurity.
In both cases, one variational parameter was used. To improve the accuracy,
two variational parameters are introduced in the present paper
in the coordinate wave functions of ground state and some exited states of divalent impurity:
\begin{align}
\label{4.1}
    {\psi _1} &= {c_1}\left| {1s,{{\vec r}_1},{\alpha _1}} \right\rangle \left| {1s,{{\vec r}_2},{\beta _1}} \right\rangle,\\
\label{4.2}
    {\psi _2} &= {c_2}\left( {\left| {1s,{{\vec r}_1},{\alpha _2}} \right\rangle \left| {1p,{{\vec r}_2},{\beta _2}} \right\rangle
    - \left| {1p,{{\vec r}_1},{\alpha _2}} \right\rangle \left| {1s,{{\vec r}_2},{\beta _2}} \right\rangle } \right),\\
\label{4.3}
    {\psi _3} &= {c_3}\left( {\left| {1s,{{\vec r}_1},{\alpha _3}} \right\rangle \left| {1p,{{\vec r}_2},{\beta _3}} \right\rangle
    + \left| {1p,{{\vec r}_1},{\alpha _3}} \right\rangle \left| {1s,{{\vec r}_2},{\beta _3}} \right\rangle } \right),\\
\label{4.4}
    {\psi _4} &= {c_4}\left( {\left| {1s,{{\vec r}_1},{\alpha _4}} \right\rangle \left| {1d,{{\vec r}_2},{\beta _4}} \right\rangle
    - \left| {1d,{{\vec r}_1},{\alpha _4}} \right\rangle \left| {1s,{{\vec r}_2},{\beta _4}} \right\rangle } \right),\\
\label{4.5}
    {\psi _5} &= {c_5}\left( {\left| {1s,{{\vec r}_1},{\alpha _5}} \right\rangle \left| {1d,{{\vec r}_2},{\beta _5}} \right\rangle
    + \left| {1d,{{\vec r}_1},{\alpha _5}} \right\rangle \left| {1s,{{\vec r}_2},{\beta _5}} \right\rangle } \right),
\end{align}
where
\begin{eqnarray}\label{5}
    \left| {j,{{\vec r}_i},{\gamma _q}} \right\rangle  &=& R_j^{}\left( {{r_i},{\gamma _q}} \right)
    Y_{{l_j}}^{{m_j}}\left( {{\theta _i},{\varphi _i}} \right) \nonumber\\
    &=& {A_j}Y_{{l_j}}^{{m_j}}\left( {{\theta _i},{\varphi _i}} \right)
    \left\{ \begin{array}{ll}
{{\bf{j}}_{{l_j}}}\left( {{k_{{n_j},{l_j}}}{r_i}} \right)\exp \left( { - {\gamma _q}{r_i}} \right), & \hbox{${r_i} \leqslant a$},\\
{{\bf{k}}_{{l_j}}}\left( {{x_{{n_j},{l_j}}}{r_i}} \right)\exp \left\{ { - {\gamma _q}\left[ {\frac{{{m_2}^*}}{{{m_1}^*}}
\left( {a - {r_i}} \right) - a} \right]} \right\}, & \hbox{${r_i} > a$},
\end{array} \right.
\end{eqnarray}
$j=1s, 1p, 1d$; index $q=1, 2, 3, 4, 5$ enumerates variational parameters for states (\ref{4.1})--(\ref{4.5});
$\gamma=\alpha, \beta$ are variational parameters, index $i=1, 2$ enumerates electrons;
$l_{1s}=0$, $l_{1p}=1$, $l_{1d}=2$; $m_{1s}=0$, $m_{1p}=-1, 0, 1$; $m_{1d}=-2, -1, 0, 1, 2$.
The spherical Bessel function of the first kind $j_\textrm{b}(z)$ and the modified spherical Bessel
function of the second kind $k_\textrm{b}(z)$ are the solutions of a Schr\"{o}dinger equation regarding the
particle in the spherical potential well with the Hamiltonian $\hat H_i^{(0)}$,
\begin{align}\label{}
    {k_{{n_j},{l_j}}} = \sqrt {\frac{{2{m_1}^*}}{{{\hbar ^2}}}E_{{n_j},{l_j}}^{(0)}} \,,
    \qquad {x_{{n_j},{l_j}}}
    = \sqrt {\frac{{2{m_2}^*}}{{{\hbar ^2}}}\left( {{U_0} - E_{{n_j},{l_j}}^{(0)}} \right)}\,,
\nonumber
\end{align}
$n_{1s}$, $n_{1p}$, $n_{1d}$ enumerates the solutions of dispersion equation when $l$ is fixed.
$A_j$ can be found from the normalization condition for the function (\ref{5}).
$\psi_1$, $\psi_3$, $\psi_5$ are functions of singlet states; $\psi_2$, $\psi_4$
are functions of triplet states. Orthogonality of total wave functions (the coordinate
part and the spin part) are provided by the orthogonality of spin parts of wave functions
and by the orthogonality of spherical harmonics. The single particle wave function ensures
the implementation of a boundary condition.

After substitution (\ref{4.1})--(\ref{4.5}) into the Schr\"{o}dinger equation with Hamiltonian (\ref{1}),
the functional was found which depends on two variational parameters for excited states and depends on
one variational parameter for the ground state. The performed procedure of numerical minimization makes it
possible to get the corresponding energy states and find the values of variational parameters,
and thus ultimately determine the wave functions.

\begin{figure}[!t]
\centerline{
\includegraphics[width=0.55\textwidth]{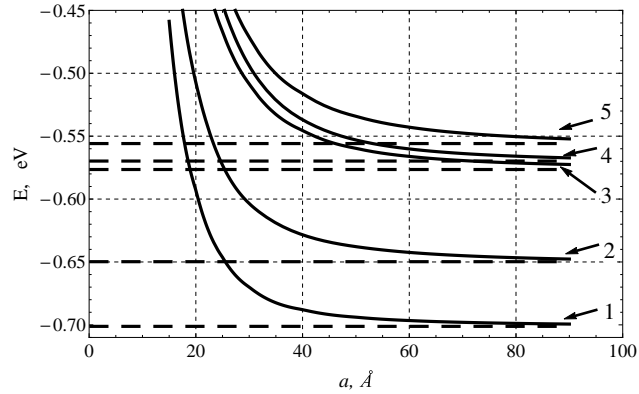}
}
\caption{The energy of divalent impurity as a function of the QD radius. Numbers denote energies
of respective states: 1~--- $\psi_1$, 2~--- $\psi_2$, 3~--- $\psi_3$, 4~--- $\psi_4$, 5~--- $\psi_5$.
Horizontal lines correspond to the energy of the divalent impurity in the bulk CdS.} \label{fig1}
\end{figure}

Calculation of electron discrete energy was performed for heterostructure CdS/SiO$_2$
with the following parameters: $m_1^* = 0.2m_0$, $m_2^* = 0.42m_0$, $\varepsilon  = \left( {5.5 + 3.9} \right)/2 = 4.7$, ${U_0} = 2.7$~eV,
where $m_0$ is free electron mass. The energy spectrum of a divalent impurity is presented in figure~\ref{fig1}.
Due to spherical symmetry, ground and excited states are degenerated by the magnetic quantum number. Figure~\ref{fig1} shows
that an increase of the QD radius leads to a decrease of the energy of the ground state which quickly becomes saturated.
For larger QD radius, the energy of excited states leads to the values corresponding to the values of the bulk crystal.
Similar dependence was observed for a monovalent impurity \cite{Boichuk1}. This dependence is
caused by a small effective Bohr radius $a_\texttt{b}^*=12.44$~{\AA}  and a large confinement.
Although the effective Bohr radius is small, the volume $a_\texttt{b}^{*3}$ approximately contains 10--12
elementary cells. This is the reason for using the Coulomb model potential interaction of electrons having an impurity.

\begin{figure}[!t]
\centerline{
\includegraphics[width=0.55\textwidth]{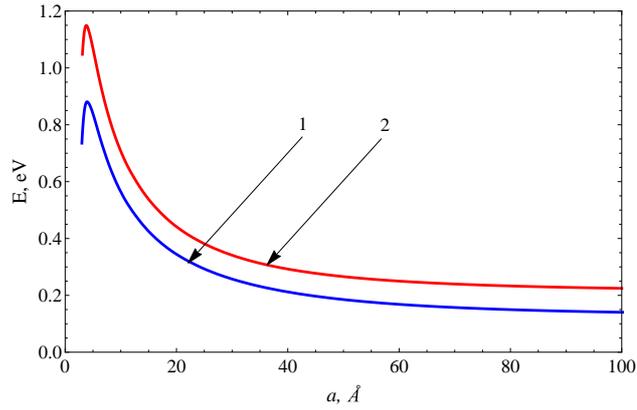}
}
\caption{(Color online) The binding energy of the univalent impurity (curve 1) and of the divalent impurity (curve 2).} \label{fig-r1}
\end{figure}

An important characteristic of the QD having a divalent or monovalent impurity is the binding energy.
In the case of a divalent impurity, $E_\textrm{b}$ is calculated by the similar formula \cite{Safwan}:
\begin{equation}\label{ionze}
    E_{\textrm{b},II}=E_0+E_{1s,Z=2}-E_1\,,
\end{equation}
where $E_0$ is the electron energy of the QD without impurities, $E_{1s,Z=2}$ is the ground state
energy of the QD having a singly ionized divalent impurity, $E_1$ is the energy of the state $\psi_1$
of the divalent impurity (\ref{4.1}). In the case of an univalent impurity, the binding energy is defined by the formula:
\begin{equation}\label{bind}
    E_{\textrm{b},I}=E_0-E_{1s,Z=1},
\end{equation}
where $E_{1s,Z=1}$ is the energy of the univalent impurity. If the QD radius reduces, the binding energy
increases in both cases. For very small radii, $E_\textrm{b}$ decreases (figure~\ref{fig-r1}). This is caused by an increase
of the probability of location of the electrons outside the QD in both cases.
However, if the QD has a divalent impurity, the binding energy is larger.

\section{Optical properties}

Energy spectrum and wave functions make it possible to calculate interlevel transitions.
Selection rules by spin variables state that transitions are possible only between
singlet-singlet and triplet-triplet states.

\begin{figure}[!b]
\centerline{
\includegraphics[width=0.55\textwidth]{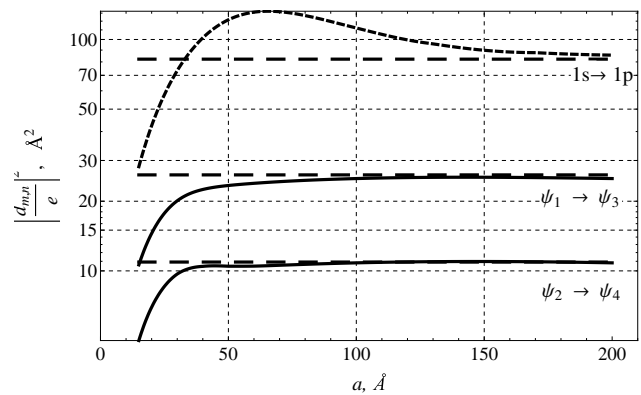}
}
\caption{The square of the dipole momentum of interlevel transitions. Solid curves
correspond to the divalent impurity. The dotted curve corresponds to the monovalent impurity.
Horizontal dashed curve denotes the square of the dipole momentum of interlevel transitions of
the monovalent and divalent impurity in the bulk crystal.} \label{fig2}
\end{figure}

Let the QD be irradiated by the linearly polarized light along the $z$ direction.
Then, in the dipole approximation, interlevel transitions are possible between
the states $\psi_1$ and $\psi_3$; $\psi_2$ and $\psi_4$; $\psi_3$ and $\psi_5$.
Dipole transition matrix elements between those states are given by:
\begin{equation}\label{6}
      {d_{13}} = \left\langle {\psi _1^{}} \right|ez\left| {\psi _3^{}} \right\rangle ,
      \qquad{d_{24}} = \left\langle {\psi _2^{}} \right|ez\left| {\psi _4^{}} \right\rangle ,
      \qquad {d_{35}} = \left\langle {\psi _3^{}} \right|ez\left| {\psi _5^{}} \right\rangle.
\end{equation}
The dependence of the square of the dipole transition matrix element on the QD radius
is presented in figure~\ref{fig2} with logarithmic scale. ${\left| {{d_{1s - 1p}}/e} \right|^2}$
for the monovalent impurity is plotted too. Figure~\ref{fig2} shows that the corresponding values for a
monovalent impurity are bigger than for the divalent one. This is due to the changes in the average distance
of electrons in their respective states. Besides, it was established that all the curves for a large QD
radii tend to the values that correspond to the values of the bulk crystal.

\begin{figure}[!t]
\centerline{
\includegraphics[width=0.55\textwidth]{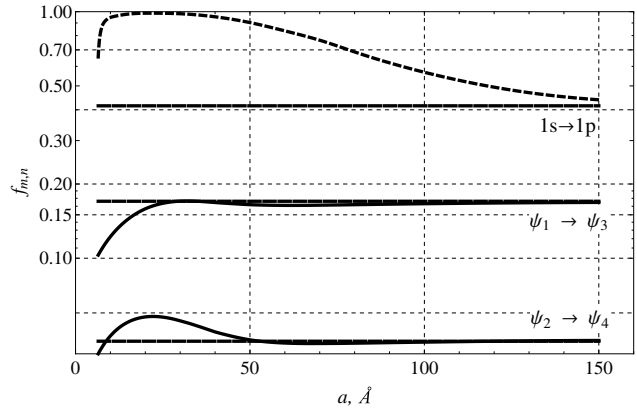}
}
\caption{The oscillator strength of interlevel transitions. Solid curves correspond
to the divalent impurity. The dotted curve corresponds to the monovalent impurity.
Horizontal curves denote the oscillator strength of interlevel transitions of the
monovalent and divalent impurity in the bulk crystal.} \label{fig3}
\end{figure}

\begin{figure}[!b]
\centerline{
\includegraphics[width=0.55\textwidth]{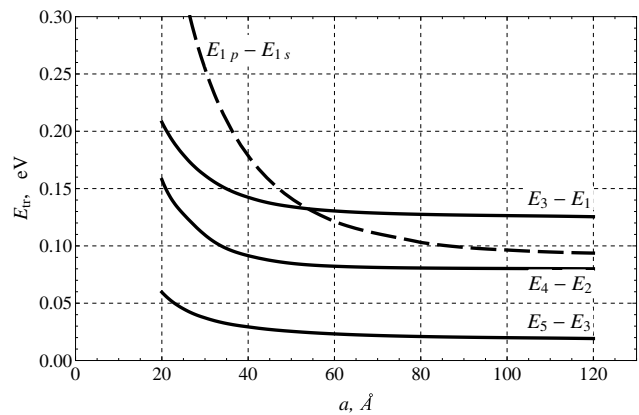}
}
\caption{The transition energy. Solid curves correspond to the divalent impurity.
Dashed curve corresponds to the monovalent impurity.} \label{fig4}
\end{figure}

The oscillator strength of interlevel transitions is also defined
\begin{equation}\label{7}
    {f_{mn}} = \frac{{2m_1^*}}{{{\hbar ^2}{e^2}}}\left( {{E_n} - {E_m}} \right){\left| {{d_{mn}}} \right|^2}.
\end{equation}
The dependences are presented in figure~\ref{fig3} with logarithmic scale.
The oscillator strength of interlevel transitions for a monovalent impurity in the
center of the QD is plotted too. This is in agreement with the result of other works
\cite{Boichuk1,Holovatsky2}. Similarly to the dipole momentum, the oscillator
strength of the divalent impurity is smaller than the oscillator strength of the
monovalent impurity. This dependence is caused by the dependence of the dipole
momentum and the transition energy $E_\textrm{tr}$=$E_n-E_m$ (figure~\ref{fig4}).

The above mentioned dependence of the dipole momentum and the transition oscillator
strength effects the height of the absorption peaks. For a two-level system,
the density matrix and iterative procedure were used to derive the absorption
coefficient \cite{Vahdani,Rezaei2,Tang}. In this approach, the linear absorption coefficient can be expressed as follows:
\begin{equation}\label{9}
    {\alpha _{m,n}}\left( \omega  \right) = \omega \sqrt {\frac{{{\mu _0}}}{{{\varepsilon _0}\varepsilon }}}
    \frac{{N{{\left| {{d_{m,n}}} \right|}^2}\hbar \Gamma }}{{{{({E_n} - {E_m} - \hbar \omega )}^2}
    + {{\left( {\hbar \Gamma } \right)}^2}}}\,,
\end{equation}
where $\varepsilon_0$ is electric constant, $\mu_0$ is magnetic constant, $c$ is
the speed of light, $N \approx 3 \cdot 10^{16}$~cm$^{-3}$ is carrier concentration,
$\hbar \Gamma$ is the scattering rate caused by the electron-phonon interaction and by
some other factors of scattering. If $\hbar \Gamma$ limits to zero, one can obtain:
\begin{equation}\label{10}
    {\alpha _{m,n}}\left( \omega  \right) = \mathop {\lim }\limits_{\hbar \Gamma  \to 0} \left( {\omega \sqrt {\frac{{{\mu _0}}}{{{\varepsilon _0}\varepsilon }}} \frac{{N{{\left| {{d_{m,n}}} \right|}^2}\hbar \Gamma }}{{{{({E_n} - {E_m} - \hbar \omega )}^2} + {{\left( {\hbar \Gamma } \right)}^2}}}} \right) = \omega \pi \sqrt {\frac{{{\mu _0}}}{{{\varepsilon _0}\varepsilon }}} N{\left| {{d_{m,n}}} \right|^2}\delta \left( {{E_n} - {E_m} - \hbar \omega } \right).
\end{equation}

In practice, sets of QD are obtained which are located on both crystal and polymer matrix or
in the solutions. Whatever method of cultivation is used, the set of QDs are always
characterized by the size dispersion. Let the QD size distribution be approximated by the Gauss function:
\begin{equation}\label{11}
    g\left( {s,\bar a,a} \right) = \frac{1}{{s\sqrt {2\pi } }}\exp \left( { - \frac{{{{\left( {a - \bar a} \right)}^2}}}{{2{s^2}}}} \right),
\end{equation}
where $a$ is the QD radius (variable), $s$ is a half-width of the distribution (\ref{11}),
which is expressed by the average radius $\bar{a}$ and the value $\sigma$ which is considered
as the variance in the QD sizes expressed in percentage: $s = \bar a\sigma /100$.
By regarding the QD dispersion (\ref{11}), the absorption coefficient is obtained for the set of QDs:
\begin{displaymath}
    {\alpha _{m,n;\textrm{system}}}\left( \omega  \right) = \omega \pi \sqrt {\frac{{{\mu _0}}}{{{\varepsilon _0}\varepsilon }}}
    N\int {g\left( {s,\bar a,a} \right)\,\,{{\left| {{d_{m,n}}\left( a \right)} \right|}^2}\delta
    \left( {{E_n}\left( a \right) - {E_m}\left( a \right) - \hbar \omega } \right)\rd a}.
\end{displaymath}
Using delta-function properties we obtain:
\begin{equation}\label{12}
    {\alpha _{m,n;\textrm{system}}}\left( \omega  \right) = \omega \pi \sqrt {\frac{{{\mu _0}}}{{{\varepsilon _0}\varepsilon }}}
    N\int {g\left( {s,\bar a,a} \right)\,\,{{\left| {{d_{m,n}}\left( a \right)} \right|}^2}\sum\limits_i
    {\frac{{\delta \left( {a - {a_{0i}}} \right)}}{{{{\left| {\frac{\rd}{{\rd a}}\left( {{E_n}\left( a \right)
    - {E_m}\left( a \right) - \hbar \omega } \right)} \right|}_{a = {a_{0i}}}}}}} \rd a},
\end{equation}
where $a_{0i}$ are simple zeros of the function $F\left( a \right) = {E_n}\left( a \right) - {E_m}\left( a \right)
- \hbar \omega$. Therefore,
\begin{equation}\label{13}
    {\alpha _{m,n;\textrm{system}}}\left( \omega  \right) = \omega \pi \sqrt {\frac{{{\mu _0}}}{{{\varepsilon _0}\varepsilon }}}
    N\sum\limits_i {\frac{{g\left( {s,\bar a,{a_{0i}}} \right)\,\,{{\left| {{d_{m,n}}
    \left( {{a_{0i}}} \right)} \right|}^2}}}{{{{\left| {\frac{\rd}{{\rd a}}\left( {{E_n}
    \left( a \right) - {E_m}\left( a \right) - \hbar \omega } \right)} \right|}_{a = {a_{0i}}}}}}}.
\end{equation}

\begin{figure}[!t]
\centerline{
\includegraphics[width=0.55\textwidth]{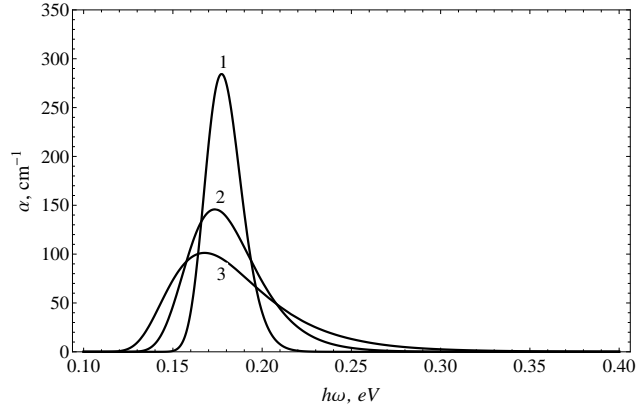}
}
\caption{The absorption coefficient of the QD system with the average radius $\bar{a}=$~40~{\AA}.
The curve 1 denotes the QD system with $\sigma=5\%$, the curve 2~--- $\sigma=10\%$, the curve 3~--- $\sigma=15\%$.} \label{fig5}
\end{figure}

\begin{figure}[!b]
\centerline{
\includegraphics[width=0.55\textwidth]{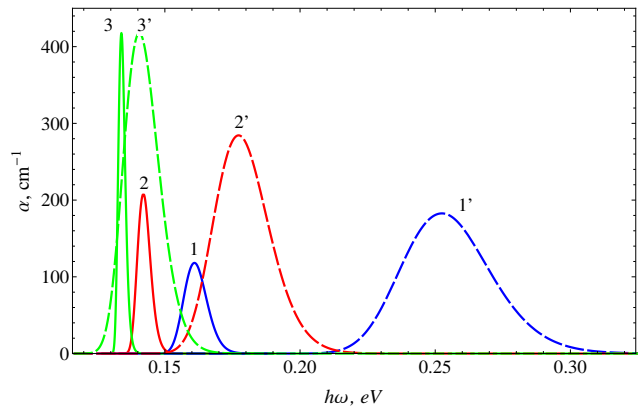}
}
\caption{(Color online) The absorption coefficient of the QD system. Solid curves 1, 2, 3 denote the
absorption coefficient of the QD with divalent impurity (transitions between singlet
states $\psi_1$, $\psi_3$), dashed curves 1', 2', 3' denote the absorption coefficient
of the QD with univalent impurity. 1, 1'~--- average radius is 30~{\AA}; 2, 2'~--- average
radius is 40~{\AA}; 3, 3'~--- average radius is 50~{\AA}.
} \label{fig6}
\end{figure}

The dependence of the absorption coefficient on the energy quant of light for different average
radii and dispersion $\sigma$ was plotted using expression (\ref{13}).

In figure~\ref{fig5} for an univalent impurity in the QD, the dependence of the QD
absorption coefficient which is caused by the $1s-1p$ transition, was plotted for three
different values of $\sigma$. The figure shows that for highly dispersed QDs, the height
of the absorption peak decreases and the absorption band blurs. This leads to an overlap
of absorption bands caused by transitions between other allowed states.
For monodispersion systems or systems with low $\sigma$, those transitions are clearly seen.
A similar situation exists for the bivalent impurity in the spherical QD. Thus, further we
consider a system of QDs with $\sigma$=5\%.

\begin{figure}[!t]
\centerline{
\includegraphics[width=0.55\textwidth]{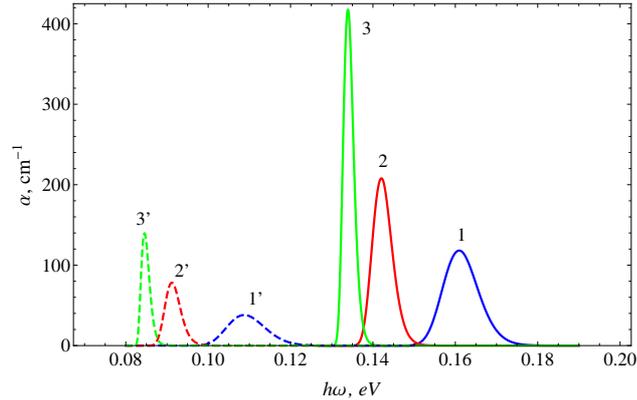}
}
\caption{(Color online) Absorption coefficient of the system of QDs with divalent impurity. Solid curves 1, 2, 3
denote absorption between singlet states $\psi_1$--$\psi_3$ , dashed curves 1', 2', 3' denote
absorption between triplet states $\psi_2$--$\psi_4$. Curves 1, 1'~--- average radius 30~{\AA}; 2,
2'~--- 40~{\AA}; 3, 3'~--- 50~{\AA}.} \label{fig7}
\end{figure}

The absorption coefficient is plotted in figure~\ref{fig6}. The figure shows that for the same
average QD radius $\bar{a}$, $1s-1p$ transition in the QD having a monovalent impurity and the respective
absorption coefficient are larger than corresponding values in the QD having a divalent impurity.
This is caused by a larger oscillator strength and dipole momentum of the interlevel transition in
the case of univalent impurity. Values of $|d_{m,n}|^2$ in the case of univalent impurity are larger,
because $\left| {\left\langle {{r_{n}}} \right\rangle  - \left\langle {{r_{m}}} \right\rangle } \right|$
is larger for the univalent impurity than for the divalent impurity. A similar explanation of the height
of absorption bands is presented in our previous works \cite{Boichuk1,Boichuk2}.
Both for the monovalent and divalent impurity, an increase of the average QD radius leads to
the shift of absorption bands into the low-energy range. When the average QD radius is less than 55~{\AA},
the absorption band caused by the transition $1s-1p$ with monovalent impurity is located in the high-energy
range in comparison with the transition $\psi_1$--$\psi_3$ of the divalent impurity. For larger $\bar{a}$,
this dependence reversed. Moreover, this can be seen in figure~\ref{fig4}.

It should be noted that the absorption of electromagnetic waves by the system of QDs having a divalent
impurity caused by the transitions between singlet states, is stronger than the corresponding
absorption between triplet states (figure~\ref{fig7}). In addition, the transition energy of triplet
states $\psi_2$, $\psi_4$ is smaller than the transition energy of singlet states $\psi_1$, $\psi_3$.
Therefore, the respective absorption bands are shifted into low-energy region. Both for the singlet-singlet
and triplet-triplet transitions, the small $\sigma$ are provided without overlapping the absorption bands which are clearly identified.

\section{Summary}

The present paper studied optical properties of the QD heterosystem CdS/SiO$_2$ having a divalent impurity in the center
of the QD, which made it possible:
\begin{itemize}
  \item to determine the energy spectrum of the QD with a divalent impurity and to show that
  this energy is smaller than the energy of the monovalent impurity in the same QD;
  \item to calculate the dipole momentum and the oscillator strength of the interlevel
  transition and to find out that the absorption between the singlet states is stronger than
  between the triplet states;
  \item to establish that the absorption bands of low dispersion systems with QD caused
  by a transition between the permitted states are clearly visible and do not overlap;
  \item to show that in the presence of a monovalent impurity, the absorption coefficient
  is larger than in the presence of a divalent impurity.
\end{itemize}

The results obtained are valid at very low temperatures. Their adjustments will be made by considering the
temperature dependence. This will be implemented in our further  works.




%
%


\ukrainianpart

\title{Міжрівневе поглинання електромагнітних хвиль нанокристалом з двовалентною домішкою}
%
%
\author{В.І. Бойчук, Р.Я. Лешко}
\address{Кафедра теоретичної фізики, Дрогобицький державний педагогічний університет ім. Івана Франка\\
вул. Стрийська, 3, 82100 Дрогобич, Львівська обл.}

\makeukrtitle

\begin{abstract}
\tolerance=3000%
У рамках методу ефективної маси обчислено спектр центральної двовалентної домішки у квантовій точці (КТ) сферичної форми. Визначено дипольні моменти та сили осциляторів міжрівневих переходів. Проаналізовано залежність лінійного коефіцієнту поглинання електромагнітних хвиль від розмірів КТ та частоти падаючої хвилі. Проведено порівняння з відповідними результатами для одновалентної домішки.
\keywords  двовалентна домішка, лінійний коефіцієнт поглинання
\end{abstract}

\end{document}